\documentclass[12pt]{article}
\usepackage{fullpage}
\usepackage{here}
\usepackage{amssymb}
\usepackage{amsfonts}
\usepackage{epsf}
\usepackage[colorlinks]{hyperref}

\begin{document}

\title{\bf Hadronic Part of the Muon {\boldmath $g-2$} \\
Estimated on the 
{\boldmath $\sigma_{\rm total}^{2003}(e^+e^-\to hadrons)$}\\ 
Evaluated Data Compilation}

\author{V.\ V.\ Ezhela, S.\ B.\ Lugovsky\ and\ O.\ V.\ Zenin\\  
 {\it Institute for High Energy Physics, Protvino, Russia}}
\date{}
\maketitle

{
\abstract{
\noindent
A comprehensive as of November 2003 and evaluated data
compilation  on $\sigma_{\mathrm tot}(e^+e^-\to hadrons)$ 
was used to estimate the lowest order hadronic contribution 
to the muon anomalous magnetic moment. The preliminary result is 
$$
a_\mu(\mathrm{had}, \mathrm{\footnotesize LO}) = 
(699.6 \pm 1.9_{\mathrm{rad}} \pm 2.0_{\mathrm{proc}} \pm 8.5_{e^+e^-})
\times 10^{-10}
$$
\noindent
The Standard Model value of the muon magnetic 
anomaly calculated by updated SM formulae published or {\tt e}-printed by 
November 2003 then reads 
$$
a_\mu = 1.165\,918\,35(87_{\mathrm{had}})(40_{\mathrm{LbL}})(03_{\mathrm{QED}})
(02_{\mathrm{EW}}) \times 10^{-3}
$$
}
}

\section*{Introduction}
Recent progress in refining the experimental and theoretical knowledge on
the muon magnetic moment anomaly, which is one of the  
most sensitive to possible new effects in particle physics 
(beyond the Standard Model) quantities, 
is summarized in the Table \ref{summary}.

\begin{table}[h]
\caption{
In the column ``Experiment'' the 1998 entry is the CODATA recommended value,
1999-2002 entries are the world averaged values quoted in the cited 
experimental papers. In the column ``Theory'' the successive theoretical values
calculated in the SM framework are presented. 
Asterisk marks the corrected result for the second 1998 and the 1999 
theoretical entries.}
\begin{center}
\begin{tabular}{|c|lc|lc|}
\hline
Year & \multicolumn{2}{|c|}{Experiment (BNL-821 et al.)} & \multicolumn{2}{|c|}{Theoretical
estimates in SM} \\
\hline
 & & & & \\[-2mm]
1998 &$1.165\,916\,02(64)\times10^{-3}$ &\cite{Mohr:2000ie}&
      $1.165\,916\,45(156)\times10^{-3}$& \cite{Alemany:1997tn}\\
     &                                  &                   & 
      $1.165\,916\,87(96)\times10^{-3}$ & \cite{Hayakawa:1997rq}\\      
1999 &$1.165\,923\,5(73)\times10^{-3}$  & \cite{Carey:dd}& 
      $1.165\,916\,3(8)\times10^{-3}$   & \cite{Hughes:fp}\\ 
2000 &$1.165\,920\,5(46)\times10^{-3}$  & \cite{Brown:2000sj}&
      $1.165\,915\,97(67)\times10^{-3}$ &\cite{Czarnecki:2000id} \\
2001 &$1.165\,920\,3(15)\times10^{-3}$  & \cite{Brown:2001mg}&
      $1.165\,918\,49(69)\times10^{-3}$ & \cite{DeTroconiz:2001wt}\\
     &                                  &
     &$1.165\,918\,56(96)\times10^{-3}$\hfill *&\cite{Hayakawa:2001bb} \\
2002 &$1.165\,920\,3(8)\times10^{-3}$   & \cite{Bennett:2002jb}&
      $1.165\,916\,93(70)(35)(04)\times10^{-3}$&\cite{Davier:2002dy} \\
\hline
\end{tabular}
\label{summary}
\end{center}
\end{table}

\noindent
It is seen that experiment is already stable and is evolving to the 
more accurate value whereas the theoretical estimates, 
in spite of intense activity, are far from being stable.

The bottleneck in the theoretical 
evaluation of $g-2$ is the  hadronic contributions, especially 
the lowest order one. It cannot be found within the perturbative approach.

The problem is traditionally treated phenomenologically: 
the imaginary part of the hadronic vacuum polarization operator 
$\Pi^{\mathrm{had}}$, through the real analyticity and asymptotic 
boundedness property can be related to the total cross section 
of the process $e^+e^- \to hadrons$ and the 
$\Pi^{\mathrm{had}}$ can be reconstructed from the experimental data 
using a dispersion relation technique. 

This requires on the one hand a complete (to date) and accurate
database of evaluated data on $\sigma_{tot}(e^+e^- \to hadrons)$ 
extracted from the original publications, and on the other hand 
a stable (traceable) and reproducible in further refinements 
method of integration of the experimental data.

The Standard Model value of the muon anomalous magnetic moment 
can be conventionally broken into following parts:

\begin{equation}
a_\mu({\scriptsize \mathrm{SM}}) = a_\mu({\scriptsize \mathrm{QED}}) +
a_\mu({\scriptsize \mathrm{EW}}) + a_\mu({\scriptsize \mathrm{had}}) 
\nonumber
\end{equation}
with
\begin{equation}
a_\mu(\mathrm{had}) = 
a_\mu(\mathrm{had}, {\scriptsize \mathrm{LO}}) +
a_\mu(\mathrm{had}, {\scriptsize \mathrm{HO}}) +
a_\mu(\mathrm{had}, {\scriptsize \mathrm{LbL}})~.
\label{eq:breakdown}
\end{equation} \\

\begin{figure}[h]
{\epsfxsize=150mm \epsfbox[-100 0 1487 510]{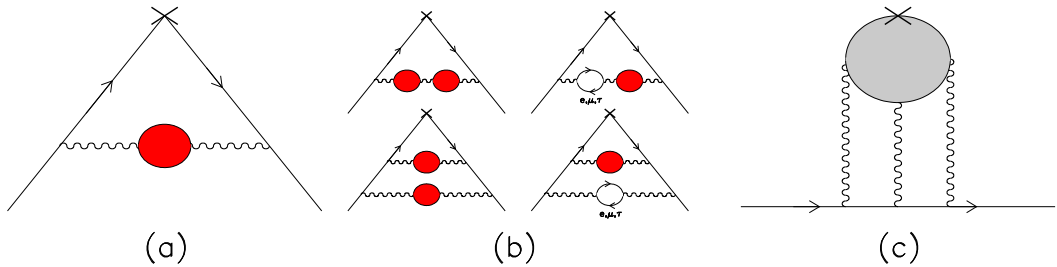}}
\caption{
Lowest {\tt (a)}, and higher order {\tt (b),(c)} 
hadronic contributions to $a_\mu$. 
Shaded and empty circles on the graphs {\tt (a),(b)} 
denote hadronic and leptonic vacuum polarization 
operators, respectively.
A shaded circle on the graph {\tt (c)} denotes an effective 
four photon vertex dominated by the pion pole. 
}\label{F1}
\end{figure}

\noindent
In the breakdown of the hadronic contribution (\ref{eq:breakdown}) 
the terms are as follows:
\begin{itemize}
\item 
$a_\mu({\scriptsize\mathrm{QED}}) = (11\,658\,470.6 \pm 0.3)\times 10^{-10}$
arises from the diagrams including only photon and 
charged lepton lines and calculated 
up to four loops (see a recent review~\cite{Knecht:2003kc} and references therein)
\item
$a_\mu({\scriptsize \mathrm{EW}}) = 
(15.4 \pm 0.1_{\mathrm{hadronic~loops}} \pm 0.2_{M_{\mathrm{Higgs}}}) 
\times 10^{-10}$
arises from one- and two-loop diagrams with $W$, $Z$ 
and Higgs internal lines \cite{Czarnecki:2002nt};
\item
$a_\mu(\mathrm{had})$ includes: 
the lowest order hadronic contribution (Fig.~\ref{F1}a)
of a typical size   
$$a_\mu(\mathrm{had}, {\scriptsize \mathrm{LO}}) \simeq 700 \times 10^{-10};$$
higher order hadronic contributions (Fig.~\ref{F1}b) \cite{Krause:1996rf}
$$a_\mu(\mathrm{had}, {\scriptsize \mathrm{HO}}) = 
(-10.1 \pm 0.6)\times 10^{-10};$$ 
light-by-light scattering contribution 
(Fig.~\ref{F1}c) (\cite{Knecht:2001qf} and references therein)
$$a_\mu(\mathrm{had}, {\scriptsize \mathrm{LbL}}) = 
(8 \pm 4) \times 10^{-10}.$$ 
\end{itemize}
 Current situation in theoretical estimates of the muon anomaly 
 is illustrated in the Table \ref{2003sum}.
\vspace*{-0.5cm}
\begin{table}[h]
\caption{Recent SM calculations using different methods and different strategies
in treatment of experimental data on $\sigma_{tot}(e^+ e^- \to hadrons)$} 
\label{2003sum}
\begin{center}
{
\begin{tabular}{|c|lc|}
\hline
 & & \\[-4mm]
 Year &\multicolumn{2}{|c|}{ Theoretical estimates in SM }\\
 \hline 
  & & \\[-4mm]
2002 &$1.165\,916\,93(70_{\mathrm{had}})(35_{\mathrm{LbL}})(04_{\mathrm{QED}+EW})\times10^{-3}$&
\cite{Davier:2002dy}\\
2002 &$1.165\,918\,89(78)\times10^{-3}        $&
\cite{Bartos:2001pg}\\
2002 &$1.165\,916\,69(74)\times10^{-3}        $&\cite{Hagiwara:2002ma}\\
2002 &$1.165\,917\,26(77)\times10^{-3}        $&\cite{Hagiwara:2002ma}\\
2003 &$1.165\,916\,96(94)\times10^{-3}        $&\cite{Jegerlehner:2003qp}\\
2003 &$1.165\,916\,75(75_{\mathrm{had}})(40_{\mathrm{LbL}})(04_{\mathrm{QED}+EW})(\times10^{-3}$
&\cite{Nyffeler:2003vb} \\
2003 &$1.165\,918\,12(127)\times10^{-3}        $&
\cite{Narison:2003ur}\\
2003 &$1.165\,918\,09(72_{\mathrm{had}})(35_{\mathrm{LbL}})(04_{\mathrm{QED}+EW})\times10^{-3}$&
\cite{Davier:2003pw}\\
2003 &$1.165\,918\,56(64_{\mathrm{had}})(35_{\mathrm{LbL}})(04_{\mathrm{QED}+EW})\times10^{-3}$&
\cite{Groote:2003kg}\\
2003 &$1.165\,918\,35(87_{\mathrm{had}})(40_{\mathrm{LbL}})(03_{\mathrm{QED}})
(02_{\mathrm{EW}}) \times 10^{-3}$& 
[{\bf this work}] \\
\hline
\end{tabular}
}
\end{center}
\end{table}

\section*{Details of the calculation}  
In this Letter we concentrate on the lowest order hadronic contribution to 
$a_\mu$.
An evaluation of the Fig.~\ref{F1}a diagram leads due to dispersion relation
to a computationally convenient representation  
of $a_\mu({\scriptsize \mathrm{had,LO}})$ \cite{K_peterman}:
\begin{equation}\label{eq:a_mu_had_disp}
a_\mu({\scriptsize \mathrm{had, LO}}) = 
4\alpha_0^2 \int^{\infty}_{m_\pi^2} 
\frac{ds}{s} K(s)\, \frac{1}{\pi}\, \mathrm{Im}\, \Pi^{\mathrm{had}}(s) = 
\frac{\alpha_0^2}{3\pi^2} \int^{\infty}_{m_\pi^2} 
\frac{ds}{s} K(s) R^{\mathrm{had}} (s)~,
\end{equation}
where 
\begin{equation}\label{eq:R}
R^{\mathrm{had}}(s) = 
\sigma_{\mathrm{tot}} (e^+e^- \to hadrons, \mathrm{lowest~order~in}~\alpha) \,
\left/  \,
\frac{4\pi\alpha_0^2}{3s} 
\right.
\end{equation}
is the well known hadronic $R$-ratio,
and the integration kernel is~\cite{K_peterman}
\begin{equation}\label{eq:K}
K(s) = \int^1_0 dx \frac{x^2 (1-x)}{x^2 + (1-x) (s/m_\mu^2)}~. 
\end{equation}
An explicit expression for the kernel at $\sqrt{s} > 2m_\mu$ reads
\begin{equation}\label{eq:K}
K(s) =
x^2 \left( 1-\frac{x^2}{2} \right) + 
(1+x)^2 \left(1 + \frac{1}{x^2} \right) 
\left(\mathrm{ln}(1+x) - x + \frac{x^2}{2} \right) 
+ \frac{1+x}{1-x} x^2 \, \mathrm{ln} x~
\end{equation}
with
$x = \left.\left(1 - \sqrt{1 - 4m_\mu^2/s}\right)\right/
\left(1 + \sqrt{1 - 4m_\mu^2/s}\right)$,
and at $\sqrt{s} < 2m_\mu$
\begin{eqnarray}\label{eq:K}
K(s) &=& \left[ -4 \left( \left(\frac{4m_\mu^2}{s} - 1 \right)^2
                           - 6 \left(\frac{4m_\mu^2}{s} - 1\right) + 1
                   \right) \mathrm{arctan}\sqrt{\frac{4m_\mu^2}{s} - 1}
         \right.           
\nonumber\\
     &         & + \sqrt{\frac{4m_\mu^2}{s} - 1}
                  \left( \left(\frac{4m_\mu^2}{s} - 1 \right)^2
                        - (3+8\mathrm{ln}2) \left(\frac{4m_\mu^2}{s} - 1 \right)
                  \right)
\nonumber\\
     &         & \left. \left.
                 + 16 \sqrt{\frac{4m_\mu^2}{s} - 1} 
                       \left(\frac{2m_\mu^2}{s} - 1 \right)\,
                   \mathrm{ln} \frac{4m_\mu^2}{s} 
                   \right] \,\, 
                   \right/ \,\, 
                   \left[ 2\,\,  \sqrt{\frac{4m_\mu^2}{s} - 1} 
                          \left( \frac{4m_\mu^2}{s} \right)^2
                   \right]~.
\nonumber
\end{eqnarray}

\subsection*{Experimental input}
Experimental data on the total cross section
of $e^+e^- \to hadrons$
were used to evaluate $R(s)$ in the range
$0.36~\mathrm{GeV} < \sqrt{s} < (12 \div 40)~\mathrm{GeV}$.
An indexed list of references to the publications related to our
evaluated data compilation of the total hadronic cross sections used  
in our analysis is given in the Appendix and can be accessed in the machine 
readable form via {\tt http://wwwppds.ihep.su:8001/eehadron.html}.
We excluded all preliminary or withdrawn by their authors data.
An overall view of the compilation is shown on the Fig.~\ref{F_R}

\begin{figure}[h]
\hspace*{-5mm}{\epsfxsize=180mm \epsfbox[0 0 567 283]{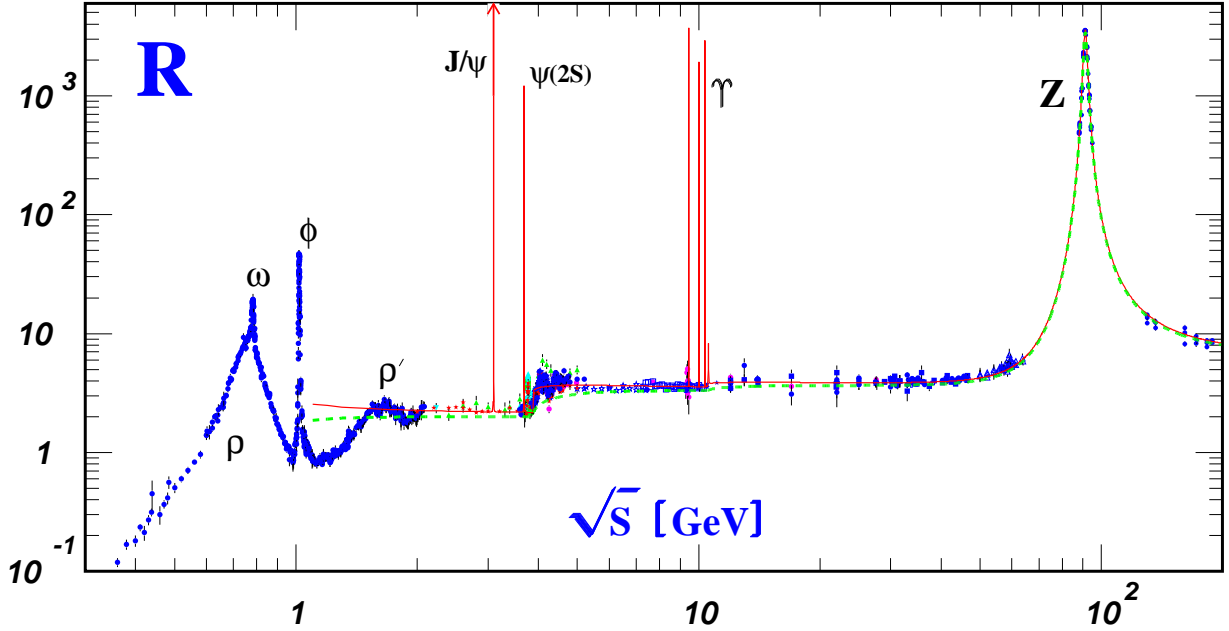}}
\caption{\footnotesize
The compilation of data on $\sigma_{\mathrm{tot}}(e^+e^-\to hadrons)$ 
rescaled to the hadronic $R$ ratio.}\label{F_R}
\end{figure}

\addtocounter{figure}{-1}

\begin{figure}
{\epsfxsize=140mm \epsfbox[-80 0 1592 850]{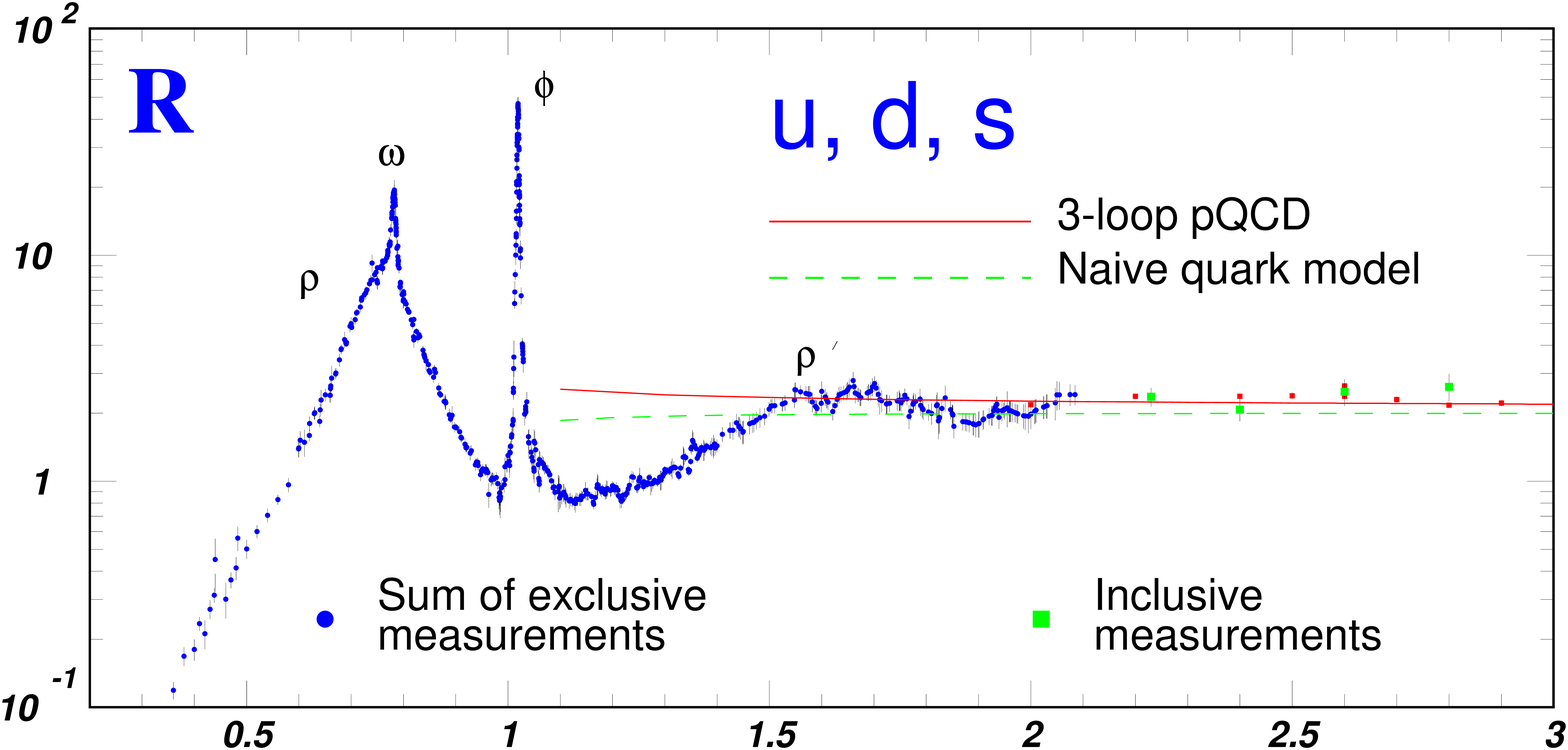}}
{\epsfxsize=140mm \epsfbox[-80 0 1621 850]{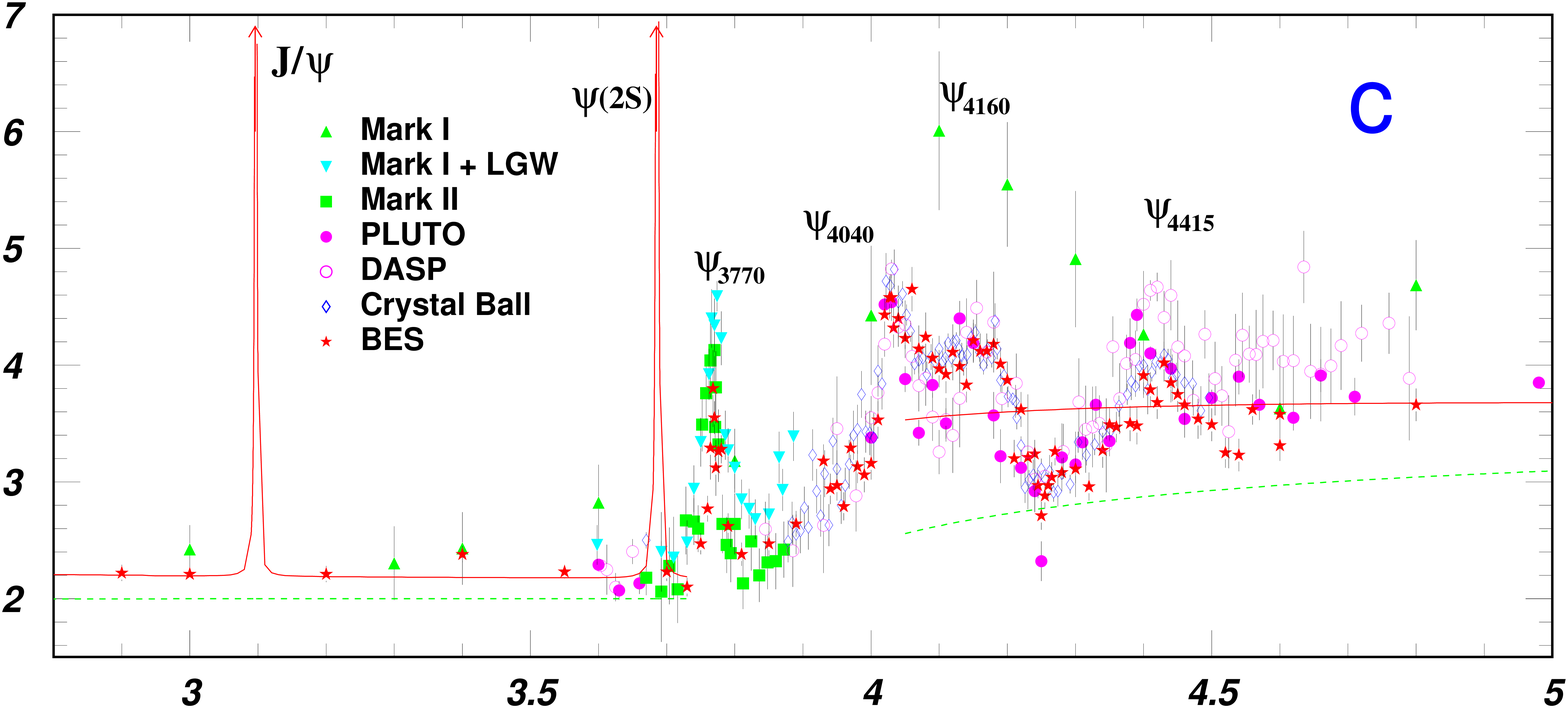}}
{\epsfxsize=140mm \epsfbox[-80 0 1621 850]{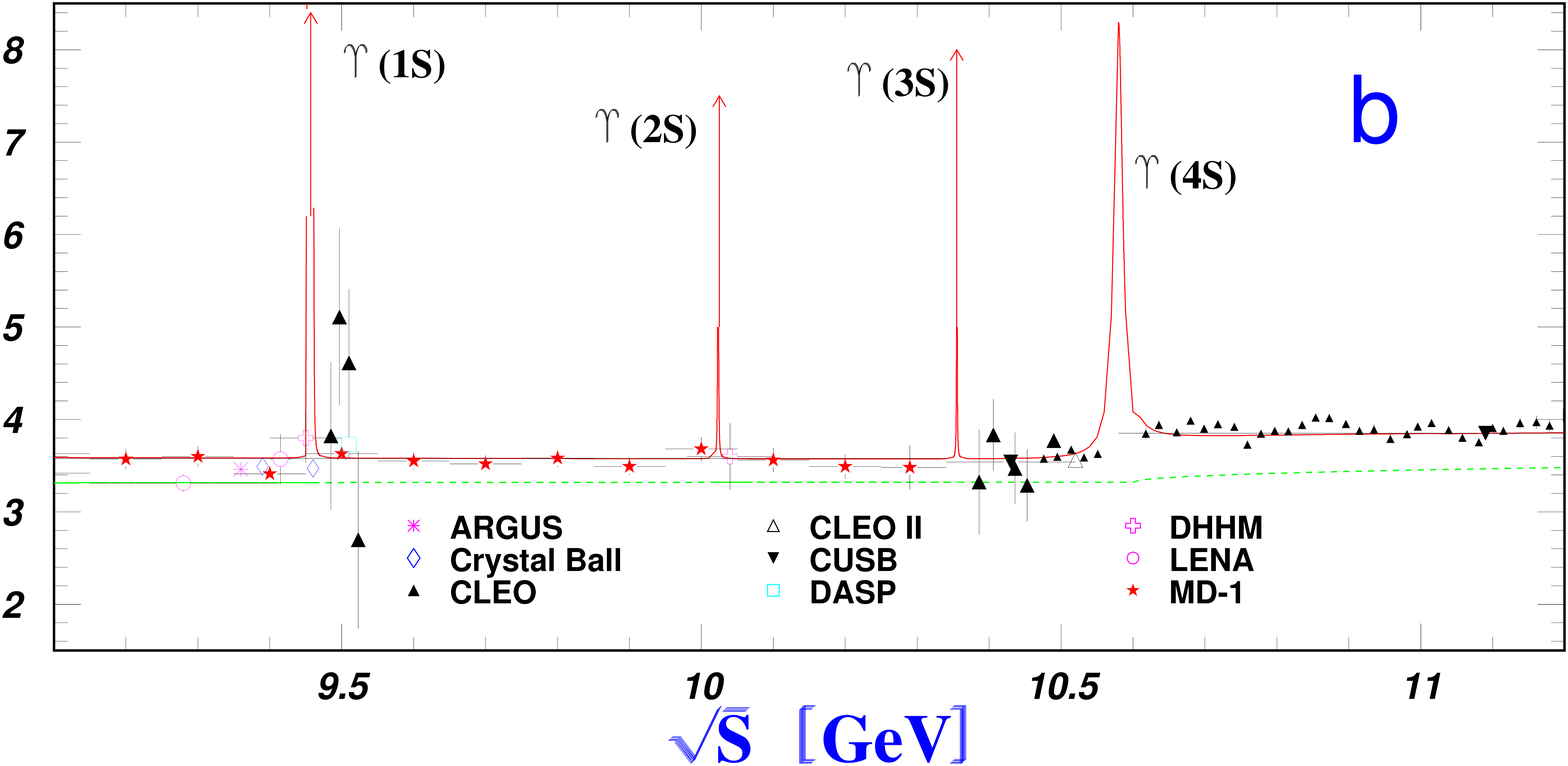}}
\caption{(continued) 
\footnotesize
Threshold regions in the $e^+e^-$ hadroproduction:
$u,d,s$-, $c$- and $b$-flavour onset.
Note the consistency between the exclusive and inclusive data
at $\sqrt{2} \sim 2$~GeV.}
\end{figure}

\noindent
{\bf Exclusive measurements.}\\[1.2ex]
{\bf At {\boldmath $\sqrt{s} < 2$}~GeV} the cross sections for 
exclusive hadronic final states were measured by 
Novosibirsk, Orsay and Frascati experiments (Tab.~\ref{Tab_excl}).
Each measurement was rescaled first to the cross section in the improved Born
approximation $\sigma^{\mathrm{IBA}}$, 
i.e. to a visible cross section corrected for ISR plus electronic vertex loops.
The latter correction is always applied in the published data, therefore 
the only things left to us were to remove partial~\cite{Bonneau:1971mk} 
or full
vacuum polarization correction if applied by the data authors
and account corrections for the photons radiated from the final state.

The partial correction for the electron vacuum polarization
usually applied in earlier  publications (before 1985)
can be removed by factor  
\begin{eqnarray}\label{eq:rad_BM}
C^{\mathrm{\sf b/m}}(s) &=& C^{\mathrm{\sf b/m}}_s(s) \times 
                       C^{\mathrm{\sf b/m}}_{\mathrm{Bhabha}}(s)~,
\nonumber\\
C^{\mathrm{\sf b/m}}_s(s) &=& \frac{1}{1 - 2\mathrm{Re}\Delta\alpha_e(s)}~,
\nonumber\\
C^{\mathrm{\sf b/m}}_{\mathrm{Bhabha}}(s) &=&
\frac{~\vspace*{10ex}
\int_{t_1}^{t_2} dt\,\, \sigma_{\mathrm{Bhabha}}
\left. (s, t)
\right\vert_{\alpha(q^2) = \frac{\alpha_0}{1-\Delta\alpha(q^2)}}
}
{~
\int_{t_1}^{t_2} dt\,\, \sigma_{\mathrm{Bhabha}}
\left. \left. (s, t)
\right. \right\vert_{\alpha(q^2) = \frac{\alpha_0}{1-\Delta\alpha_e(q^2)}}
}
\end{eqnarray}
Here $C^{\mathrm{\sf b/m}}_s(s)$ removes the correction for 
the electronic vacuum polarization in the $s$-channel and
$C^{\mathrm{\sf b/m}}_{\mathrm{Bhabha}}(s)$
properly corrects the 
large angle Bhabha scattering cross section
used as a luminosity monitor in the low energy experiments.
The squared momentum transfer range $[t_1,t_2]$ is individual for each 
detector~\cite{Swartz:1995hc}. 
Finally, $\Delta\alpha_e(s)$ and $\Delta\alpha(s)$ are 
related to the electronic and full vacuum polarization operators as 
\begin{equation}
\Delta\alpha_e(s) = \Pi_{e}(s)/s~~,~~
\Delta\alpha(s) = \Pi(s)/s~,
\end{equation}
respectively.
An expression for the one loop electronic (generally leptonic)
vacuum polarization operator $\Pi_{\ell}(s)$ can be found 
elsewhere (see, e.g.~\cite{Landau_Lifshitz}).
The hadronic part of vacuum polarization cannot be calculated perturbatively
and is related to the experimental hadronic $R$ ratio as
\begin{equation}\label{eq:da_had}
\Delta\alpha_{\mathrm{had}}(s) = -\frac{\alpha_0}{3\pi}\, s\, 
\int_{m_\pi^2}^\infty \frac{R(s')\, ds'}{s'(s' - s - i0)}~.
\end{equation}

The full vacuum polarization correction applied by some later
experiments is removed by the factor
\begin{equation}\label{eq:rad_full}
C^{\mathrm{full}}(s) = \frac{1}{1 - 2\mathrm{Re}\Delta\alpha(s)}~,
\end{equation}
where 
\begin{equation}
\Delta\alpha(s) = \sum_{\ell = e,\mu,\tau} \Delta\alpha_{\ell}(s) + 
                  \Delta\alpha_{\mathrm{had}}(s)
\end{equation}
with $\Delta\alpha_{\mathrm{had}}(s)$ calculated 
according to~(\ref{eq:da_had}).

Turning off doubtful radiative corrections allows to estimate 
the ``radiative'' uncertainty 
of~$a_\mu(\mathrm{had}, {\scriptsize \mathrm{LO}})$.
             
After rescaling the data to $\sigma^{\mathrm{IBA}}$
their weighed average $\overline{\sigma}^{\mathrm{IBA}}(s)$ 
for each exclusive hadronic channel is found by the inimization 
of the bilinear form
\begin{equation}\label{eq:chi2_excl}
Q = \sum_{(k)} (C^{(k)})^{-1}_{ij} 
    \left( \overline{\sigma}^{\mathrm{IBA}}(s^{(k)}_i) - 
               {\sigma^{\mathrm{IBA}}}^{(k)}_i
    \right)
    \left( \overline{\sigma}^{\mathrm{IBA}}(s^{(k)}_j) - 
               {\sigma^{\mathrm{IBA}}}^{(k)}_j
    \right)~.
\end{equation}
Here an index $k$ runs over the experiments measured the given channel,
indices $i$ and $j$ enumerate the data points within the $k$-th experiment
and $C^{(k)}$ is the error matrix of the $k$-th experiment.
We conservatively
assumed no correlations between systematic uncertainties of data coming
from different publications even if the  measurements were performed at
the same facility. 
Unfortunately, publications (especially older ones) 
do not provide enough information to
split a total systematic uncertainty into separate sources 
which is necessary to find correlations between the experiments. 
Indeed, 100\% correlation might be among 
uncertainties coming from a machine luminosity determination,
radiative corrections, {\it etc.}, i.e. from the procedures common for all
experiments.  
{\it Post factum} crude estimates~\cite{Davier:2002dy} 
based on expert judgements
are also questionable because the lack of descriptions of the 
systematic error sources can easily lead to double counting. 
These considerations justify our refusal (for the moment) 
to take into account the correlations between experiments 
despite a possible overestimate of the $e^+e^-$ uncertainty 
of $a_\mu$. This point requires further investigation.

Next, the averaged cross sections of each exclusive final state
are summed up to give the total cross section 
$\sigma_{\mathrm{tot}}^{\mathrm{IBA}}(e^+e^- \to hadrons)$.
The latter divided by $4\pi\alpha^2(s)/3s$ gives the desired hadronic 
ratio $R(s)$ entering the integral~(\ref{eq:a_mu_had_disp}).

Summing the contributions of exclusive hadronic final states
is not always straightforward.
To account for the missing hadronic final states
the isospin symmetry relations are exploited where possible (see 
the discussion in~\cite{Davier:2002dy}). 

%
%

Note that determination of $\Delta\alpha(s)$ in 
Eqs.~(\ref{eq:rad_BM}),~(\ref{eq:rad_full}) 
in turn requires to evaluate the dispersion relation integral~(\ref{eq:da_had})
also containing the $R$ ratio. 
For this reason an iterative procedure was applied:
at zero iteration $\Delta\alpha(s)$ was calculated in the naive approximation
including one loop QED contributions from all quarks and charged leptons
plus contributions from $\phi(1020)$, $J/\psi$, $\psi(2S)$ and 
$\Upsilon(1S)\dots\Upsilon(4S)$ resonances;
then the obtained naive $\Delta\alpha(s)$ was used to properly rescale
the experimental data to the $R$ ratio which in turn
was used for the evaluation of $\Delta\alpha(s)$ 
to be utilized at the next iteration.
This process proved to be well convergent even in the vicinities of 
narrow resonances which contribution to $\Delta\alpha(s)$ was 
obtained analytically as will be explained later.
It turns out that two iterations are enough with the current level of 
accuracy.\\

\noindent
{\bf {\boldmath $\pi^+\pi^-$, $\pi^0\gamma$} threshold regions.}\\

\noindent
The lowest experimental point of our compilation 
lies at $\sqrt{s} = 0.36$~GeV, well
above $\pi^+\pi^-$ and $\pi^0\gamma$ production thresholds.
The hadronic cross section in the $2m_\pi < \sqrt{s} < 0.36$~GeV  
range can be evaluated using the chiral perturbation theory (ChPT)
parametrization of the pionic formfactor
\begin{equation}\label{eq:F_ChPT}
F^{\mathrm{ChPT}}_\pi(s) = 1 + \frac{{\scriptsize<}r^2{\scriptsize>}_\pi}{6} s + c_1 s^2 + c_2 s^3 
+ {\cal O}(s^4)~, 
\end{equation}
where the squared pionic charge radius 
$<r^2>_\pi = (11.27 \pm 0.21)$~GeV$^{-2}$ follows from the fit of 
space-like data~\cite{Amendolia:1986wj} and the parameters $c_1$, $c_2$
are fitted to the time-like data in the range 
$2m_\pi < \sqrt{s} < 0.6$~GeV~\cite{Davier:2002dy}.

The two pion production cross section above the threshold then reads 
\begin{equation}\label{eq:sig_ChPT}
\sigma_{\pi^+\pi^-} = \frac{\pi|\, \alpha(s)|^2}{3s}\, (1 - 4m_\pi^2/s)^{3/2}\, 
                      |F^{\mathrm{ChPT}}_\pi (s)|^2 \,\, .
\end{equation}
The $\pi^0\gamma$ cross section in the range $m_\pi < \sqrt{s} < 0.6$~GeV is 
much smaller and parametrized using the $\pi^0-\gamma^*\gamma$ transition
formfactor~\cite{Knecht:2001qf}.
The phenomenological parametrization of the low energy hadronic cross section
can be reliably used up to $\sqrt{s_{\mathrm{ChPT}}} = 0.5$~GeV.
The uncertainty due to the variation of $\sqrt{s_{\mathrm{ChPT}}}$
in the range 
0.36~--~0.6~GeV is folded to the procedural error 
of~$a_\mu(\mathrm{had}, {\scriptsize \mathrm{LO}})$. \\

\noindent
{\bf Inclusive measurements.}\\

\noindent
At $\sqrt{s} \ge 2$~GeV the $e^+e^-$
experiments mostly measure the total cross section   
of an inclusive production of hadrons (Tab.~\ref{Tab_incl}).
Data published after 1978 seem to be fully corrected for ISR, 
electronic vertex loops and vacuum polarization.
Earlier data with only electronic vacuum polarization correction need to be
properly rescaled as mentioned above (Eq.~\ref{eq:rad_BM}).
Data rescaled to the correct $R$ ratio were weighed in the 
Eq.~(\ref{eq:chi2_excl}) manner to give the averaged $R$ ratio entering 
into dispersion relation integrals~(\ref{eq:a_mu_had_disp}) and (\ref{eq:da_had}).\\

\noindent
{\bf Resonances.} \\[1.2ex]
The contributions of
$\omega(782)$, $\phi(1020)$, $\psi(3770)$, $\psi(4040)$ and $\psi(4160)$
resonances are already contained in the cross section data
as their widths are larger than a typical machine energy spread. 
Although,  in this preliminary work
we account for 
the contribution of a relatively broad $\phi(1020)$ meson
using the relativistic Breit-Wigner parametrization
\begin{equation}\label{eq:R_res}
R_{res} (s) \, = \, \sigma_{\mathrm{BW}}(s) 
              \, \left/\,  \frac{4\pi|\alpha(s)|^2}{3s} \right. \,
            = \, \frac{9}{|\alpha(s)|^2} \,  
              \frac{s\, \Gamma_{ee}\Gamma \, \frac{s}{M^2}}
                   {(s-M)^2 + s\, \Gamma^2} \,\, ,
\end{equation}
where $\Gamma_{ee}$ and $\Gamma$ are physical electronic and total widths
of the resonance given by PDG~\cite{Hagiwara:fs}. 
Narrow $J/\psi$, $\psi(2S)$ and
$\Upsilon(nS), n=1..4$ resonances were treated in the same way.

Note that a dispersion relation in the form~(\ref{eq:da_had})
should not be used
to find the contribution of a narrow resonance to the running 
 $\alpha(s)$ as the $R$ ratio~(\ref{eq:R_res}) in turn   
contains $\alpha(s)$ rapidly varying in the vicinity of a resonance.
Instead, we use another form of the dispersion integral relating 
the values of $\alpha(s)$ with and without 
the contribution of the resonance:
\begin{eqnarray}\label{eq:res_disp}
\alpha(s)\, - \, \alpha_{\mathrm{without}}(s) 
&=& -\frac{s}{4\pi^2} \int \frac{\sigma_{\mathrm{BW}}(s')\,\, ds'}{s' - s - i0}
\nonumber \\
&=& -\frac{s}{\pi} 
      \int \frac{3\, \Gamma_{ee}\Gamma \, \frac{s'}{M^2}}
                {(s'-M^2)^2 +s'\Gamma^2} \,
           \frac{ds'}{s' - s - i0}\,\,.  
\end{eqnarray}
This expression can be easily obtained 
from the analytic properties of the re-summed photon propagator.\\

\noindent
{\bf High energy tail}\\

\noindent
The $R$ ratio at $\sqrt{s} > 12$~GeV
can be reliably evaluated using the perturbative QCD. 
We used a three loop pQCD approximation 
taking into account the effect of quark masses~\cite{Zenin:2001yx}.
A variation of the lower boundary of pQCD usage 
in the range 12~--~40~GeV results in a negligible additional uncertainty of 
$a_\mu(\mathrm{had,LO})$.

\section*{Discussion}
 
To cross-check the obtained value of the lowest order hadronic contribution 
to the muon magnetic anomaly
\begin{equation}\label{eq:this}
a_\mu(\mathrm{had}, \mathrm{\footnotesize LO}) = 
(699.6 \pm 1.9_{\mathrm{rad}} \pm 2.0_{\mathrm{proc}} \pm 8.5_{e^+e^-})
\times 10^{-10}\,\, ,
\end{equation}
we repeated the procedure on the subset
of our compilation of $\sigma_{\mathrm{tot}}(e^+e^-\to hadrons)$ data
used in recent papers~\cite{Davier:2002dy,Davier:2003pw}.
The result 
\begin{equation}\label{eq:cross_eidelman}
a_\mu(\mathrm{had}, \mathrm{\footnotesize LO}, \mathrm{\footnotesize subset}) = 
(694.5 \pm 1.9_{\mathrm{rad}} \pm 2.0_{\mathrm{proc}} \pm 8.8_{e^+e^-})
\times 10^{-10}~,
\end{equation}
is consistent with 
the corrected~\cite{Davier:2003pw} $e^+e^-$ based result of the 
paper~\cite{Davier:2002dy}
\begin{equation}\label{eq:eidelman}
a_\mu(\mathrm{had }, \mathrm{\footnotesize LO}, \cite{Davier:2003pw}) = 
(696.3 \pm 3.6_{\mathrm{rad+proc}} \pm 6.2_{e^+e^-})\times 10^{-10} \,\, .
\end{equation}
The uncertainty induced by the $e^+e^-$ data is larger in our work 
because of significant differences in the data treatment:
\begin{description}
\item[different integration procedures:]
we obtain the total $R$ ratio first 
and then integrate it without averaging within small energy bins each
including several experimental points;
on the other hand, in~\cite{Davier:2002dy,Davier:2003pw} 
the contributions of each hadronic final state 
were added separately and the aforementioned energy averaging was applied;
\item[different treatment of systematic errors:] 
no correlations 
between different experiments in this work versus significant correlations
nominated in~\cite{Davier:2002dy,Davier:2003pw}.
\end{description}
These items will be clarified in the forthcoming publication.

\section*{Conclusion}
A new estimate of the lowest order hadronic contribution to the muon
anomalous magnetic moment was obtained using a comprehensive (as of
November 2003) compilation of evaluated data on total hadronic cross 
sections in $e^+e^-$ collisions.
The preliminary result that is free of any extra admissions on the data
not documented in the original experimental publications reads

$$
a_\mu(\mathrm{had}, \mathrm{\footnotesize LO}) = 
(699.6 \pm 1.9_{\mathrm{rad}} \pm 2.0_{\mathrm{proc}} \pm 8.5_{e^+e^-})
\times 10^{-10}\,\, ,
$$
where the first error is due to the uncertainty in the radiative corrections 
to the $e^+e^-$ data, the second one is procedural and the last one
is due to the experimental errors of the $e^+e^-$ data.
The value of the muon magnetic anomaly then reads as
$$
a_\mu = (11 659 183.5
         \pm 8.7_{\mathrm{had}} 
         \pm 4.0_{\mathrm{LbL}}
         \pm 0.3_{\mathrm{QED}}
         \pm 0.2_{\mathrm{EW}}
        ) \times 10^{-10}
\,\, ,
$$
where the errors are from the hadronic, light-by-light scattering, 
pure QED and electroweak contributions, respectively.
This result deviates from
the experimental ``world average'' of~$a_\mu$~\cite{Sichtermann:2003cc}~by
$$
(-19.5 \pm 9.6_{\mathrm{theor}} \pm 8.0_{\mathrm{exp}}) \times 10^{-10}\,\, ,
$$
i.e.\ at $\sim 1.5\sigma$ level.     

As it can be seen from the Table \ref{2003sum} our estimate is well matched with all other 2002-2003 estimates based on the $e^+\, e^- \to hadrons$ data. The differences are due to slightly different databases used
and different methods of incorporating experimental data into final estimates.
It seems that to make a further progress in the refinement of these
estimates it will be useful to standardize the database to meet all 
aspects of the scientific database quality: completeness, accuracy and 
traceability of the data transference from original publications to 
the evaluated data compilations, transparency of the data evaluation 
procedures, and easy access to the evaluated database in computer 
readable form
for physics and education communities. Some  steps towards 
such compilations were undertaken by the COMPAS and HEPDATA groups
under auspices of the PDG 
collaboration~\cite{Hagiwara:fs,Zenin:2001yx,Whalley:2003qr}.

The standardized and maintained evaluated data compilation will allow
to join efforts and find most stable and reliable method of incorporating
experimental data into the theoretical estimates of the hadronic 
contributions to the high precision observables and to trace the 
consistency of the different experimental evidences with the 
theoretical estimates in the Standard Model.

\vspace*{2ex}
\paragraph*{Acknowledgments.} 
This work was supported in part by the Russian Foundation of Basic 
Research under grant RFBR-01-07-90392.  
The authors are grateful to M.\ R.\ Whalley 
for his contributions to the IHEP\ {\tt PPDS(ReacData)}\  and
{\tt PPDS(CrossSections)} databases. We thank our colleagues 
R.~M.~Barnett, D.~E.~Groom, R.~Miquel, and A.~M.~Zaitsev for fruitful 
discussions and encouragements. We also thank Z.~G.~Zhao and Y.~S.~Zhu 
for providing us with the BES numerical data.

%


\def\cs{$\sigma$}
\def\bm{\sf \sf b/m}
\def\ful{\sf full}
\def\isr{\sf ISR}
\def\aco{ACO}
\def\bcf{ADONE-BCF}
\def\cmddva{VEPP-2M-CMD-2}
\def\cmd{VEPP-2M-CMD}
\def\acodmraz{ACO-DM1}
\def\dmraz{DCI-DM1}
\def\dmdva{DCI-DM2}
\def\fen{ADONE-FENICE}
\def\gg{ADONE-$\gamma\gamma{}2$}
\def\mtrin{DCI-M3N}
\def\mea{ADONE-MEA}
\def\mupi{ADONE-$\mu\pi$}
\def\nd{VEPP-2M-ND}
\def\olya{VEPP-2M-OLYA}
\def\snd{VEPP-2M-SND}
\def\tof{VEPP-2M-TOF} 
\def\vepp{VEPP-2}
%
%
\def\pip{$\pi^+$} 
\def\pim{$\pi^-$}
\def\piz{$\pi^0$} 
\def\kp{$K^+$}
\def\km{$K^-$}
\def\ks{$K_S$}
\def\kl{$K_L$}
\def\pbar{$\bar{p}$}
\def\nbar{$\bar{n}$}
%
%
\def\EPJ#1#2#3#4{Eur. Phys. J. {\bf #1} (#3) #2 & #4}
\def\LNC#1#2#3#4{Nuovo Cim. Lett. {\bf #1} (#3) #2 & #4}
\def\NC#1#2#3#4{Nuovo Cim. {\bf #1} (#3) #2 & #4}
\def\NP#1#2#3#4{Nucl. Phys. {\bf #1} (#3) #2 & #4}
\def\NPPS#1#2#3#4{Nucl. Phys. Proc. Suppl. {\bf #1} (#3) #2 & #4}
\def\PL#1#2#3#4{Phys. Lett. {\bf #1{}B} (#3) #2 & #4}
\def\PR#1#2#3#4{Phys. Rev. {\bf #1} (#3) #2 & #4}
\def\PRPL#1#2#3#4{Phys. Rep. {\bf #1} (#3) #2 & #4}
\def\SJNP#1#2#3#4{Sov. J. Nucl. Phys. {\bf #1} (#3) #2 & #4}
\def\YF#1#2#3#4{Yad. Fiz. {\bf #1} (#3) #2 & #4}
\def\ZETFP#1#2#3#4{Zh. Exp. Th. Fiz. Pisma {\bf #1} (#3) #2 & #4}
\def\ZP#1#2#3#4{Z. Phys. {\bf #1} (#3) #2 & #4}
\def\PREP#1#2#3{#1 (#2) & #3}
%
\def\EX#1#2#3#4{#1 & #4 & #2 & #3}
%


\vspace*{5ex}

\section*{Appendix}

Column titles in Tabs.~\ref{Tab_excl},~\ref{Tab_incl} 
are self-explaining. 
$R$ and $\sigma$ in the ``Obs.'' column denote the types of observables
in the original papers: $R$ ratio and cross section, respectively. 
The abbreviations in the last column of Tab.~\ref{Tab_excl} and in the fifth
column of Tab.~\ref{Tab_incl}
denote the types of radiative corrections (RC)
applied in the original publications:
\begin{description}
\item[{\sf b/m}] -- data corrected for the initial state radiation (ISR),
$e^+e^-$ vertex loops and electronic vacuum 
polarization~~\cite{Bonneau:1971mk};
\item[{\sf full}] -- data corrected for ISR, 
$e^+e^-$ vertex loops and full vacuum polarization;
\item[{\sf ISR}] -- data corrected for ISR and $e^+e^-$ vertex loops only.
\end{description}
Data not used in the calculations are marked by an asterisk 
(to compare, see the recent papers~\cite{Davier:2002dy,Davier:2003pw,Whalley:2003qr} 
where the extended experimental bibliography is also presented).


\def\flg{\hspace*{-2.ex}$^*$\ }

\begin{table}[h]
\boldmath
\small
\caption{Exclusive hadronic cross section measurements.}\label{Tab_excl}
\vspace*{2ex}
\footnotesize
\begin{tabular}{ l 
                 l 
                 l 
                 c 
                 c 
               }
\cline{1-5}
\\[-1ex]               
Experiment & Reference & Author {\it et al.}  & Obs. & RC
\\[1ex]
\cline{1-5}

\\[1.5ex]\multicolumn{5}{c}{ \pip \pim  ($\gamma$)}

\\[1.5ex] \EX{\vepp    }{ \cs }{ \bm  }{ \flg \PL{41}{205}{1972}{Balakin, V.E.} }  
\\[1ex] \EX{ \olya    }{ \cs }{ \bm  }{ \NP{B256}{365}{1985}{Barkov, L.M.} }  
\\[1ex] \EX{ \tof     }{ \cs }{ \bm  }{ \SJNP{33}{368}{1980}{Vasserman, I.B.} }  
\\[1ex] \EX{ \cmd     }{ \cs }{ \bm  }{ \NP{B256}{365}{1985}{Barkov, L.M.} }  
\\[1ex] \EX{ \aco     }{ \cs }{ \bm  }{ \flg \PL{39}{289}{1972}{Benaksas, D.} }
\\[1ex] \EX{ \dmraz   }{ \cs }{ \bm  }{ \PL{76}{512}{1978}{Quenzer, A.} }  
\\[1ex] \EX{ \dmdva   }{ \cs }{ \ful }{ \PL{220}{321}{1989}{Bisello, D.} }  
\\[1ex] \EX{ \mea     }{ \cs }{ \bm  }{ \LNC{28}{337}{1980}{Esposito, B.} }  
\\[1ex] \EX{ \bcf     }{ \cs }{ \bm  }{ \LNC{15}{393}{1976}{Bollini, D.} }  
\\[1ex] \EX{ \cmddva  }{ \cs }{ \ful }{ \PL{527}{161}{2002}{Akhmetshin, R.R.} }  
\\[1ex] \EX{          }{     }{      }{ Erratum \PREP{\flg hep-ex/0308008}{2003}{Akhmetshin, R.R.} }   

\\[2ex]\multicolumn{5}{c}{ \pip \pim \piz }

\\[1.5ex] \EX{\cmddva  }{ \cs }{\isr  }{ \PL{476}{33}{2000}{Akhmetshin, R.R.} }  
\\[1ex] \EX{          }{     }{      }{ Erratum \PREP{\flg hep-ex/0308008}{2003}{Akhmetshin, R.R.} }   
\\[1ex] \EX{          }{ \cs }{\isr  }{ \PL{364}{199}{1995}{Akhmetshin, R.R.} }  
\\[1ex] \EX{          }{ \cs }{\isr  }{ \PL{434}{426}{1998}{Akhmetshin, R.R.} }  
\\[1ex] \EX{ \snd     }{ \cs }{\isr  }{ \PR{D66}{032001}{2002}{Achasov, M.N.} }  
\\[1ex] \EX{          }{ \cs }{\isr  }{ \PR{D63}{072002}{2001}{Achasov, M.N.} }  
\\[1ex] \EX{          }{ \cs }{\isr  }{ \PR{D68}{052006}{2003}{Achasov, M.N.} }  
\\[1ex] \EX{ \nd      }{ \cs }{\isr  }{ \PRPL{202}{99}{1991}{Dolinsky, S.I.} }  
\\[1ex] \EX{ \cmd     }{ \cs }{\isr  }{ \PREP{Novosibirsk 89-15}{1989}{Barkov, L.M.} }  
\\[1ex] \EX{ \dmraz   }{ \cs }{\isr  }{ \NP{B172}{13}{1980}{Cordier, A.} }  
\\[1ex] \EX{ \dmdva   }{ \cs }{\isr  }{ \ZP{C56}{15}{1992}{Antonelli, A.} }  
\\[1ex] \EX{ \aco     }{ \cs }{ \bm  }{ \flg \PL{28}{513}{1968}{Augustin, J.E.} }  
\\[1ex] \EX{          }{ \cs }{ \bm  }{ \flg \PL{42}{507}{1972}{Benaksas, D.} } 
\\[1ex] \EX{ \gg      }{ \cs }{ \bm  }{  \NP{B184}{31}{1981}{Bacci, C.} } 
\\[1ex] \EX{ \mea     }{ \cs }{ \bm  }{  \LNC{28}{195}{1980}{Esposito, B.} }  

\\[2ex] \cline{1-5} \end{tabular}
\end{table}

\addtocounter{table}{-1}


\begin{table}
\boldmath
\small
\caption{(continued) Exclusive hadronic cross section measurements.}
\vspace*{2ex}
\footnotesize
\begin{tabular}{ l 
                 l 
                 l 
                 c 
                 c 
               }
\cline{1-5}
\\[-1ex]               
Experiment & Reference & Author {\it et al.}  & Obs. & RC
\\[1ex]
\cline{1-5}

\\[2ex]\multicolumn{5}{c}{ \piz $\gamma$ }

\\[1.5ex] \EX{\snd     }{ \cs }{\isr  }{ \EPJ{C12}{25}{2000}{Achasov, M.N.} }  
\\[1ex] \EX{          }{ \cs }{\isr  }{  \PL{559}{171}{2003}{Achasov, M.N.} }  
\\[1ex] \EX{ \aco     }{ \cs }{\isr  }{  \PL{63}{352}{1976}{Cosme, G.} } 

\\[1.5ex]\multicolumn{5}{c}{ $\eta$ $\gamma$ }

\\[1.5ex] \EX{\cmddva  }{ \cs }{\isr  }{ \PL{364}{199}{1995}{Akhmetshin, R.R.} }  
\\[1ex] \EX{          }{ \cs }{\isr  }{ \PL{509}{217}{2001}{Akhmetshin, R.R.} }    
\\[1ex] \EX{ \snd     }{ \cs }{\isr  }{ \EPJ{C12}{25}{2000}{Achasov, M.N.} }  
\\[1ex] \EX{ \aco     }{ \cs }{\isr ?}{ \PL{63}{352}{1976}{Cosme, G.} }  

\\[1.5ex] \multicolumn{5}{c}{ $\omega$ $<$ \piz $\gamma$ $>$ \piz }

\\[1.5ex] \EX{\snd     }{ \cs }{\isr  }{ \PL{486}{29}{2000}{Achasov, M.N.} }  
\\[1ex] \EX{          }{ \cs }{\isr  }{ \NP{B569}{158}{2000}{Achasov, M.N.} }  
\\[1ex] \EX{ \cmddva  }{ \cs }{\isr  }{ \PL{562}{173}{2003}{Akhmetshin, R.R.} }  
\\[1ex] \EX{ \nd      }{ \cs }{\isr  }{ \PL{174}{453}{1986}{Dolinsky, S.I.} }  
\\[1ex] \EX{ \dmdva   }{ \cs }{\isr  }{ \PREP{LAL-90-35}{1990}{Bisello, D.} } 
 
\\[2ex]\multicolumn{5}{c}{ \pip \pim  $2$\piz }

\\[1.5ex] \EX{\olya    }{ \cs }{\isr  }{ \ZETFP{43}{497}{1986}{Kurdadze, L.M.} }  
\\[1ex] \EX{ \nd      }{ \cs }{\isr  }{ \PRPL{202}{99}{1991}{Dolinsky, S.I.} }  
\\[1ex] \EX{ \cmddva  }{ \cs }{\isr  }{ \PL{466}{392}{1999}{Akhmetshin, R.R.} }  
\\[1ex] \EX{ \snd     }{ \cs }{\isr  }{ \PREP{Budker~INP 2001-34}{2001}{Achasov, M.N.} }  
\\[1ex] \EX{ \mtrin   }{ \cs }{ \bm  }{ \NP{B152}{215}{1979}{Cosme, G.} }  
\\[1ex] \EX{ \gg      }{ \cs }{ \bm  }{ \NP{B184}{31}{1981}{Bacci, C.} }  
\\[1ex] \EX{ \aco     }{ \cs }{ \bm  }{ \PL{63}{349}{1976}{Cosme, G.} }  
\\[1ex] \EX{ \dmdva   }{ \cs }{\isr  }{ \PREP{LAL-90-35}{1990}{Bisello, D.} }  
\\[1ex] \EX{ \mea     }{ \cs }{ \bm  }{ \LNC{31}{445}{1981}{Esposito, B.} }

\\[2ex] \cline{1-5} \end{tabular}
\end{table}

\addtocounter{table}{-1}

\newpage

\begin{table}
\boldmath
\small
\caption{(continued) Exclusive hadronic cross section measurements.}
\vspace*{2ex}
\footnotesize
\begin{tabular}{ l 
                 l 
                 l 
                 c 
                 c 
               }
\cline{1-5}
\\[-1ex]               
Experiment & Reference & Author {\it et al.}  & Obs. & RC
\\[1ex]
\cline{1-5}

\\[2ex]\multicolumn{5}{c}{ $2$\pip $2$\pim }

\\[1.5ex] \EX{\olya    }{ \cs }{\isr  }{ \ZETFP{47}{432}{1988}{Kurdadze, L.M.} }  
\\[1ex] \EX{ \cmd     }{ \cs }{\isr  }{ \SJNP{47}{248}{1988}{Barkov, L.M.} }  
\\[1ex] \EX{ \cmddva  }{ \cs }{\isr  }{ \PL{475}{190}{2000}{Akhmetshin, R.R.} }  
\\[1ex] \EX{          }{ \cs }{\isr  }{ \PL{491}{81}{2000}{Akhmetshin, R.R.} }  
\\[1ex] \EX{          }{ \cs }{\isr  }{ \PL{466}{392}{1999}{Akhmetshin, R.R.} }  
\\[1ex] \EX{ \snd     }{ \cs }{\isr  }{ \PREP{Budker~INP 2001-34}{2001}{Achasov, M.N.} }  
\\[1ex] \EX{ \nd      }{ \cs }{\isr  }{ \PRPL{202}{99}{1991}{Dolinsky, S.I.} }  
\\[1ex] \EX{ \mtrin   }{ \cs }{ \bm  }{ \NP{B152}{215}{1979}{Cosme, G.} }  
\\[1ex] \EX{ \dmraz   }{ \cs }{ \bm  }{ \PL{81}{389}{1979}{Cordier, A.} }  
\\[1ex] \EX{          }{ \cs }{ \bm  }{ \PL{109}{129}{1982}{Cordier, A.} }  
\\[1ex] \EX{ \dmdva   }{ \cs }{\isr  }{ \PREP{LAL-90-35}{1990}{Bisello, D.} }  
\\[1ex] \EX{ \mea     }{ \cs }{ \bm  }{ \LNC{28}{195}{1980}{Esposito, B.} }  
\\[1ex] \EX{ \mupi    }{ \cs }{ \bm  }{ \NC{13A}{593}{1973}{Grilli, M.} }  
\\[1ex] \EX{ \gg      }{ \cs }{ \bm  }{ \PL{95}{139}{1980}{Bacci, C.} }  
\\[1ex] \EX{ \aco     }{ \cs }{ \bm  }{ \PL{63}{349}{1976}{Cosme, G.} }

\\[1.5ex] \multicolumn{5}{c}{ $2$\pip $2$\pim \piz }

\\[1.5ex] \EX{\cmd    }{ \cs }{\isr  }{ \SJNP{47}{248}{1988}{Barkov, L.M.} }  
\\[1ex] \EX{ \mtrin   }{ \cs }{ \bm  }{ \NP{B152}{215}{1979}{Cosme, G.} }  
\\[1ex] \EX{ \gg      }{ \cs }{ \bm  }{ \NP{B184}{31}{1981}{Bacci, C.} }  
\\[1ex] \EX{ \mea     }{ \cs }{ \bm  }{ \LNC{31}{445}{1981}{Esposito, B.} }  
\\[1ex] \EX{ \mupi    }{ \cs }{ \bm  }{ \NC{13A}{593}{1973}{Grilli, M.} }  

\\[1.5ex]\multicolumn{5}{c}{ $\omega$ $<$ \pip \pim  \piz $>$ \pip \pim  }

\\[1.5ex] \EX{\dmraz   }{ \cs }{ \bm  }{ \PL{106}{155}{1981}{Cordier, A.} }   
\\[1ex] \EX{ \dmdva   }{ \cs }{ \bm  }{ \ZP{C56}{15}{1992}{Antonelli, A.} }  
\\[1ex] \EX{ \cmddva  }{ \cs }{\isr  }{ \PL{489}{125}{2000}{Akhmetshin, R.R.} } 

\\[2ex] \multicolumn{5}{c}{ $\eta$ \pip \pim  }

\\[1.5ex] \EX{\cmddva  }{ \cs }{\isr  }{ \PL{489}{125}{2000}{Akhmetshin, R.R.} } 
\\[1ex] \EX{ \nd      }{ \cs }{\isr  }{ \PRPL{202}{99}{1991}{Dolinsky, S.I.} }   

\\[2ex]\multicolumn{5}{c}{ \pip \pim  $3$\piz }

\\[1.5ex] \EX{\mtrin  }{ \cs }{ \bm  }{ \NP{B152}{215}{1979}{Cosme, G.} }  
\\[1ex] \EX{ \mea     }{ \cs }{ \bm  }{ \LNC{25}{5}{1979}{Esposito, B.} }

\\[2ex] \cline{1-5} \end{tabular}
\end{table}

\addtocounter{table}{-1}

\newpage

\begin{table}
\boldmath
\small
\caption{(continued) Exclusive hadronic cross section measurements.}
\vspace*{2ex}
\footnotesize
\begin{tabular}{ l 
                 l 
                 l 
                 c 
                 c 
               }
\cline{1-5}
\\[-1ex]               
Experiment & Reference & Author {\it et al.}  & Obs. & RC
\\[1ex]
\cline{1-5}

\\[2ex]\multicolumn{5}{c}{ $3$\pip $3$\pim }

\\[1.5ex] \EX{\cmd     }{ \cs }{\isr  }{ \SJNP{47}{248}{1988}{Barkov, L.M.} }  
\\[1ex] \EX{ \dmraz   }{ \cs }{ \bm  }{ \PL{107}{145}{1981}{Bisello, D.} } 
\\[1ex] \EX{ \dmdva   }{ \cs }{\isr  }{ \PREP{Roma U. thesis}{1986}{Schioppa, M.} }
\\[1ex] \EX{ \mtrin   }{ \cs }{ \bm  }{ \NP{B152}{215}{1979}{Cosme, G.} }  
\\[1ex] \EX{ \mupi    }{ \cs }{ \bm  }{ \NC{13A}{593}{1973}{Grilli, M.} }  

\\[2ex]\multicolumn{5}{c}{ $2$\pip $2$\piz $2$\pim }

\\[1.5ex] \EX{\cmd     }{ \cs }{\isr  }{ \SJNP{47}{248}{1988}{Barkov, L.M.} }  
\\[1ex] \EX{ \mea     }{ \cs }{ \bm  }{ \LNC{31}{445}{1981}{Esposito, B.} } 
\\[1ex] \EX{ \mtrin   }{ \cs }{ \bm  }{ \NP{B152}{215}{1979}{Cosme, G.} }  
\\[1ex] \EX{ \gg      }{ \cs }{ \bm  }{ \NP{B184}{31}{1981}{Bacci, C.} }  
\\[1ex] \EX{ \mupi    }{ \cs }{ \bm  }{ \NC{13A}{593}{1973}{Grilli, M.} }  
\\[1ex] \EX{ \dmdva   }{ \cs }{\isr  }{ \PREP{Roma U. thesis}{1986}{Schioppa, M.} }

\\[1.5ex] \multicolumn{5}{c}{ \kp \km }

\\[1.5ex] \EX{\vepp    }{ \cs }{ \bm  }{\PL{41}{205}{1972}{Balakin, V.E.} }  
\\[1ex] \EX{ \olya    }{ \cs }{\isr  }{ \PL{107}{297}{1981}{Ivanov, P.M.} }  
\\[1ex] \EX{ \bcf     }{ \cs }{ \bm  }{ \PL{46}{261}{1973}{Bernardini, M.} }  
\\[1ex] \EX{ \acodmraz}{ \cs }{ \bm  }{ \PREP{LAL-80-35}{1980}{Grosdidier, G.} }   
\\[1ex] \EX{          }{ \cs }{ \bm  }{ \PL{99}{257}{1981}{Delcourt, B.} }
\\[1ex] \EX{ \dmdva   }{ \cs }{\isr  }{ \ZP{C39}{13}{1988}{Bisello, D.} } 
\\[1ex] \EX{ \mea     }{ \cs }{ \bm  }{ \LNC{28}{337}{1980}{Esposito, B.} }  
\\[1ex] \EX{ \cmd     }{ \cs }{\isr  }{ \PREP{Novosibirsk 83-85}{1983}{Anikin, G.V.} } 
\\[1ex] \EX{ \cmddva  }{ \cs }{\isr  }{ \PL{364}{199}{1995}{Akhmetshin, R.R.} }  
\\[1ex] \EX{ \snd     }{ \cs }{\isr  }{\PR{D63}{072002}{2001}{Achasov, M.N.} }  

\\[2ex] \cline{1-5} \end{tabular}
\end{table}

\addtocounter{table}{-1}

\newpage

\begin{table}
\boldmath
\small
\caption{(continued) Exclusive hadronic cross section measurements.}
\vspace*{2ex}
\footnotesize
\begin{tabular}{ l 
                 l 
                 l 
                 c 
                 c 
               }
\cline{1-5}
\\[-1ex]               
Experiment & Reference & Author {\it et al.}  & Obs. & RC
\\[1ex]
\cline{1-5}

\\[2ex] \multicolumn{5}{c}{ \ks \kl }

\\[1.5ex] \EX{\dmraz   }{ \cs }{ \bm  }{ \PL{99}{261}{1981}{Mane, F.} } 
\\[1ex] \EX{ \olya    }{ \cs }{ \bm  }{  \ZETFP{36}{91}{1982}{Ivanov, P.M.} } 
\\[1ex] \EX{ \cmd     }{ \cs }{ \bm  }{  \PREP{Novosibirsk 83-85}{1983}{Anikin, G.V.} } 
\\[1ex] \EX{ \cmddva  }{ \cs }{\isr  }{ \PL{364}{199}{1995}{Akhmetshin, R.R.} }  
\\[1ex] \EX{          }{ \cs }{\isr  }{ \PL{551}{27}{2003}{Akhmetshin, R.R.} }  
\\[1ex] \EX{          }{ \cs }{\isr  }{ \PL{466}{385,}{1999}{Akhmetshin, R.R. } }
\\[1ex] \EX{          }{     }{      }{ Errata {\it ibid.} {\bf 508B} (2001) 217,}
\\[1ex] \EX{          }{     }{      }{ ~~~~~~~~~~ hep-ex/0308008 (2003) }
\\[1ex] \EX{ \snd     }{ \cs }{\isr  }{\PR{D63}{072002}{2001}{Achasov, M.N.} }  

\\[1.5ex]\multicolumn{5}{c}{ \kp \km \piz }

\\[1.5ex] \EX{\dmdva  }{ \cs }{\isr  }{ \NPPS{21}{111}{1991}{Bisello, D.} } 
\\[1ex] \EX{          }{ \cs }{\isr  }{ \ZP{C52}{227}{1991}{Bisello, D.} }  

\\[2ex]\multicolumn{5}{c}{ \ks \kp \pim  $+$ \ks \km \pip }

\\[1.5ex] \EX{\dmraz   }{ \cs }{ \bm  }{ \PL{112}{178}{1982}{Mane, F.} }  
\\[1ex] \EX{ \dmdva   }{ \cs }{\isr  }{ \ZP{C52}{227}{1991}{Bisello, D.} }  

\\[1.5ex]\multicolumn{5}{c}{ \kp \km \pip \pim  }

\\[1.5ex] \EX{\dmraz   }{ \cs }{ \bm  }{ \PL{110}{335}{1982}{Cordier, A.} }  
\\[1ex] \EX{ \dmdva   }{ \cs }{\isr  }{ \NPPS{21}{111}{1991}{Bisello, D.} }

\\[2ex] \multicolumn{5}{c}{ \ks$+X$ }
\\[1.5ex] \EX{\dmraz   }{ \cs }{ \bm  }{ \PREP{LAL-82-46}{1982}{Mane, F.} }  

\\[2ex]\multicolumn{5}{c}{ $p$ \pbar}

\\[1.5ex] \EX{\dmraz    }{ \cs }{ \bm }{ \PL{86}{395}{1979}{Delcourt, B.} }  
\\[1ex] \EX{ \dmdva    }{ \cs }{ \bm }{ \NP{B224}{379}{1983}{Bisello, D.} }  
\\[1ex] \EX{ \fen      }{ \cs }{ \ful}{ \NP{B517}{3}{1998}{Antonelli, A.} }  

\\[2ex]\multicolumn{5}{c}{ $n$ \nbar }

\\[1.5ex] \EX{ \fen      }{ \cs }{ \ful}{ \NP{B517}{3}{1998}{Antonelli, A.} }  

\\[2ex] \cline{1-5} \end{tabular}
\end{table}


\def\ful{\sf full}
\def\bm{\sf b/m}
\def\cs{$\sigma$}

\def\EX#1#2#3#4{#4 & #1 & #2 & #3  \\}
\def\JO#1#2#3#4#5{#4-#5 & #1~(#2) & #3 } 
\def\PL#1#2#3#4#5#6{#5-#6 & Phys.~Lett.~{\bf #1{}B} (#3) #2 & #4  }
\def\PR#1#2#3#4#5#6{#5-#6 & Phys.~Rev.~{\bf #1} (#3) #2 & #4 }
\def\PRL#1#2#3#4#5#6{#5-#6 & Phys.~Rev.~Lett.~{\bf #1} (#3) #2 & #4 }
\def\PRPL#1#2#3#4#5#6{#5-#6 & Phys.~Rep.~{\bf #1} (#3) #2 & #4 }
\def\ZP#1#2#3#4#5#6{#5-#6 & Z.~Phys.~{\bf #1} (#3) #2 & #4 }

\begin{table}
\caption{Inclusive hadronic cross section measurements.}\label{Tab_incl}
\vspace*{3ex}
\scriptsize
\begin{tabular}{ l 
                 l 
                 l 
                 c 
                 c 
                 l 
                }
\cline{1-6}
\\[-0.7ex]
Experiment & 
Reference  & 
 Author {\it et al.} & 
Obs.       & 
RC & 
${\mathrm E}_{\mathrm{cm}}$ $[$GeV$]$
\\[1ex]
\cline{1-6}
\\[1ex]
\EX{\cs    }{\bm  }{2.23            }{\flg \PL{58}{478}{1975}{Bartoli, B.}{ADONE}{MEA}                                                       }
\EX{R      }{\ful }{2.0 - 4.8          }{\PRL{88}{101802}{2002}{Bai, J.Z.}{BEPC}{BES} }
\EX{R      }{\ful }{2.6 - 5.0          }{\PRL{84}{594}{2000}{Bai, J.Z.}{BEPC}{BES} }
\EX{\cs    }{\bm  }{2.4 - 5.0          }{\flg \PRL{34}{764}{1975}{Augustin, J.E.}{SPEAR}{SMAG~$^{\dagger}$}                                                  }
\EX{R      }{\ful }{3.598 - 3.886          }{\flg \PRL{39}{526}{1977}{Rapidis, P.A.}{SPEAR}{SMAG+LGW}                                               }
\EX{R      }{\ful }{3.670 - 4.496          }{\JO{SLAC-PUB-4160}{1986}{Osterheld, A.}{SPEAR}{Crystal Ball}                                              }
\EX{R      }{\ful }{5.0 - 7.4          }{\JO{SLAC-PUB-5160}{1989}{Edwards, C.}{SPEAR}{Crystal Ball}                                           }
\EX{R      }{\ful }{3.670 - 3.872          }{\JO{SLAC-219}{1979}{Schindler, R.H.}{SLAC}{MARK-II}                                                    }

\EX{R      }{\bm  }{3.6025 - 5.1950          }{\PL{76}{361}{1978}{Brandelik, R.}{DORIS}{DASP}}

\EX{R      }{\ful }{7.440 - 9.415          }{\ZP{C15}{299}{1982}{Niczyporuk, B.}{DORIS-II}{LENA} }
\EX{R      }{\ful }{9.360               }{\flg \ZP{C54}{13}{1992}{Albrecht, H.}{DORIS-II}{ARGUS} }
\EX{R      }{\ful }{9.39 - 9.46          }{\ZP{C40}{49}{1988}{Jakubowski, Z.}{DORIS-II}{Crystal Ball} }
\EX{R      }{\ful }{9.45 - 10.04         }{\flg \ZP{C6}{125}{1980}{Bock, P.}{DORIS-II}{DHHM} }
\EX{R      }{\ful }{9.51            }{\PL{116}{383}{1982}{Albrecht, H.}{DORIS-II}{DASP}                                                 }

\EX{R      }{\ful }{7.30 - 10.29         }{\ZP{C70}{31}{1996}{Blinov, A.E.}{VEPP-4}{MD1}                                                    }

\EX{R      }{\ful }{10.43 - 11.09        }{\PRL{48}{906}{1982}{Rice, E.}{CESR}{CUSB}                                                              }
\EX{R      }{\ful }{10.49           }{\PR{D29}{1285}{1984}{Giles, R.}{CESR}{CLEO} } 
\EX{R      }{NO   }{10.60 - 11.20   }{\PRL{54}{381}{1985}{Besson, D.}{CESR}{CLEO~$^{\dagger\dagger}$} } 
\EX{R      }{\ful }{10.52           }{\PR{D57}{1350}{1998}{Ammar, R.}{CESR}{CLEO~II} } 

\EX{R      }{\ful }{3.6 - 30.8         }{\PRPL{83}{151}{1982}{Criegee, L.}{DORIS/PETRA}{PLUTO}                                       }
\EX{R      }{\ful }{12.0 - 41.4        }{\flg \ZP{C22}{307}{1984}{Althoff, M.}{PETRA}{TASSO} }

\EX{R      }{\ful }{12.00 - 31.25        }{\flg \ZP{C4}{87}{1980}{Brandelik, R.}{PETRA}{TASSO} }
\EX{R      }{\ful }{14.03 - 43.70        }{\flg \ZP{C47}{187}{1990}{Braunschweig, W.}{PETRA}{TASSO} }
\EX{R      }{\ful }{41.45 - 44.20       }{\PL{138}{441}{1984}{Althoff, M.}{PETRA}{TASSO} }

\EX{R      }{\ful }{12.00 - 46.47        }{\PRPL{148}{67}{1987}{Naroska, B.}{PETRA}{JADE}}
\EX{R      }{\ful }{12.00 - 46.47        }{\PR{D34}{681}{1986}{Adeva, B.}{PETRA}{MARK-J} }
\EX{R      }{\ful }{31.57          }{\flg \PL{85}{463}{1979}{Barber, D.P.}{PETRA}{MARK-J} }
\EX{R      }{\ful }{34.85         }{\flg \PL{108}{63}{1982}{Barber, D.P.}{PETRA}{MARK-J}                                                   }

\EX{R      }{\ful }{14.0 - 46.6        }{\PL{183}{400}{1987}{Behrend, H.J.}{PETRA}{CELLO} }

\EX{R      }{\ful }{29.0           }{\PR{D31}{1537}{1985}{Fernandez, E.}{PEP}{MAC}                                                     }
\EX{R      }{\ful }{29.0            }{\flg \PR{D43}{34}{1991}{von Zanthier, C.}{PEP}{MARK-II} }

\EX{R      }{\ful }{50.0 - 61.4        }{\flg \PR{D42}{1339}{1990}{Kumita, T.}{TRISTAN}{AMY} }
\EX{R      }{\ful }{50.0 - 61.4        }{\flg \PL{234}{525}{1990}{Adachi, I.}{TRISTAN}{TOPAZ} }
\EX{\cs    }{\ful }{57.77          }{\flg \PL{347}{171}{1995}{Miyabayashi, K.}{TRISTAN}{TOPAZ} }
\EX{\cs    }{\ful }{57.37 - 59.84        }{\flg \PL{304}{373}{1993}{Abe, T.}{TRISTAN}{TOPAZ} }
\EX{R      }{\ful }{50.0 - 52.0        }{\flg \PL{198}{570}{1987}{Yoshida, H.}{TRISTAN}{VENUS} }
\EX{R      }{\ful }{63.6 - 64.0        }{\flg \PL{246}{297}{1990}{Abe, K.}{TRISTAN}{VENUS} }
\\[-0.5ex]
\cline{1-6}
\end{tabular}
~\\[2ex]
$^{\dagger}$ \footnotesize{The measured cross section 
excludes all-neutral final states and not used in our analysis.
}\\[0.5ex]
$^{\dagger\dagger}$ \footnotesize{Data were not corrected for ISR and 
thus not used  in our analysis.}\\[1ex]

All preliminary data are excluded. 
Data covering $J/\psi$, $\Psi(2S)$ and $\Upsilon(1S)..\Upsilon(4S)$
are also omitted as they are 
distorted by machine energy spread and 
cannot be directly used for the evaluation of
$\Delta\alpha_{\mathrm{QED}}^{\mathrm{had}}(s)$ and 
$a_\mu(\mathrm{had},\mathrm{LO})$.

Data from the SPEAR-DELCO experiment 
$[$Phys.~Rev.~Lett.~{\bf 40}~(1978)~671$]$ are excluded as they were 
not corrected for the $\tau^+\tau^-$ contamination of the hadronic sample. 
\end{table}

\end{document}